\documentclass[11pt]{article}
\title{A Heuristic Approach for Treating Pathologies of Truncated Sum Rules in Limit Theory of Nonlinear Susceptibilities}
\author{Mark G. Kuzyk  \\ Department  of Physics and Astronomy, Washington State University \\ Pullman, WA  99164-2814  \\ kuz@wsu.edu}
\usepackage{graphics}
\usepackage{amsmath}
\begin{document}
\maketitle

\begin{abstract}
The Thomas Kuhn Reich sum rules and the sum-over-states (SOS) expression for the hyperpolarizabilities are truncated when calculating the fundamental limits of nonlinear susceptibilities.  Truncation of the SOS expression can lead to an accurate approximation of the first and second hyperpolarizabilities due to energy denominators, which can make the truncated series converge to within 10\% of the full series after only a few excited states are included in the sum.  The terms in the sum rule series, however, are weighted by the state energies, so convergence of the series requires that the position matrix elements scale at most in inverse proportion to the square root of the energy.  Even if the convergence condition is met, serious pathologies arise, including self inconsistent sum rules and equations that contradict reality.  As a result, using the truncated sum rules alone leads to pathologies that make any rigorous calculations impossible, let alone yielding even good approximations.  This paper discusses conditions under which pathologies can be swept under the rug and how the theory of limits, when properly culled and extrapolated using heuristic arguments, can lead to a semi-rigorous theory that successfully predicts the behavior of all known quantum systems, both when tested against exact calculations or measurements of broad classes of molecules.

\end{abstract}

\section{Introduction}

Almost a decade and a half ago, the fundamental limits of the nonlinear-optical susceptibilities were calculated.\cite{kuzyk00.01,kuzyk00.02} This theory has determined an upper bound of the hyperpolarizability and second hyperpolarizability that has survived the test of time, where hundreds and perhaps even thousands of molecules have all been shown to fall short of the limits.\cite{kuzyk13.01}

Clays was the first to use the scaling properties of the calculated limits of the first hyperpolarizability to show how hyper-Rayleigh scattering measurements that do not account for two-photon fluorescence appear to exceed the {\em apparent limit},\cite{clays01.01} which is about a factor of 30 below the fundamental limit.\cite{kuzyk03.01,kuzyk03.02}  When the fluorescence background is properly taken into account, the measurements fall below the apparent limit.  These same scaling properties have been used to understand why most molecules fall bellow the factor of 30 gap\cite{tripa04.01,perez07.02} and to identify the new paradigm of modulated conjugation as a way to break this `artificial' barrier.\cite{perez07.01,perez09.01} Sum rules, scaling, and fundamental limit theory have thus been used as a tool for understanding the nonlinear optical response of molecules.\cite{perez09.02}

Though the limits seem to hold for all systems,\cite{kuzyk13.01} are a useful tool for comparing molecules of disparate shapes and sizes,\cite{kuzyk13.01} have been used for making better molecules through a new paradigm,\cite{zhou06.01,perez07.01,perez09.01} and are useful in studying the effects of geometry\cite{kuzyk06.02,shafe12.01} and topology\cite{lytel13.01} on the nonlinear optical response, many questions about the fundamental basis of limit theory arise.  These include the validity of the assumption that any quantum system with a hyperpolarizability at the fundamental limit is represented by a three-level model.  Though the three-level ansatz, as it is called, seems to hold up in numerical calculations, \cite{zhou07.02,ather12.01,burke13.01} it has not been rigourously proven.  Indeed, using the same approach as was originally used in getting the limits, it can be shown in a special case to lead to a divergence, known as the many-state catastrophe.\cite{shafe13.01}

The goal of this paper is to reconcile these issues and to determine how the sum rules are to be used and interpreted in nonlinear-optical calculations.

\section{Background}

Here we review the approach for arriving at the fundamental limits using the sum rules. Each of the assumptions are stated and their implications discussed.

The sum rules for a system with $N_{el}$ electrons are given by,
\begin{equation}\label{SumRules}
\sum_{n=0}^\infty \xi_{mn} \xi_{np} \left( e_n - \frac {1} {2} \left[ e_m + e_p \right] \right) = \delta_{m,p},
\end{equation}
where the ground state energy has been subtracted from each energy, then normalized such that $e_n = (E_n - E_0)/(E_1 - E_0)$; and, the matrix elements of the position operators, $x_{ij}$, are normalized to the maximum value,
\begin{equation}\label{xMAX}
x_{MAX} = \sqrt{ \frac { \hbar^2 N_{el}} {2 m_e E_{10}}},
\end{equation}
where $E_{10} = E_1 - E_0$, to get $\xi_{ij} = x_{ij} / x_{MAX}$.

For simplicity, we also assume that the dipole matrix is real.  The three-level model then gives the six independent sum rules,
\begin{eqnarray}\label{3L sum rules}
(m,p) = (0,0):&&  e_1 \left| \xi_{01}\right|^2 + e_2 \left| \xi_{02}\right|^2 = e_1  \label{(0,0)} \\
(m,p) = (1,1):&&  e_{21} \left| \xi_{12}\right|^2  = e_1 \left( 1 +  \left| \xi_{01}\right|^2 \right) \label{(1,1)} \\
(m,p) = (0,1):&&  e_1 \xi_{01} \left(\xi_{11} - \xi_{00} \right) + \left(e_{20} + e_{21} \right) \xi_{12} \xi_{02} = 0  \label{(0,1)} \\
(m,p) = (0,2):&&  e_2 \xi_{02} \left(\xi_{22} - \xi_{00} \right) =  \left(e_{21} - e_1 \right) \xi_{01} \xi_{12} \label{(0,2)} \\
(m,p) = (1,2):&&  e_{21} \xi_{12} \left(\xi_{22} - \xi_{11} \right) = \left(e_1 + e_2 \right) \xi_{01} \xi_{02} \label{(1,2)} \\
(m,p) = (2,2):&&  e_2 \left| \xi_{02}\right|^2 + e_{21} \left| \xi_{12}\right|^2 = - e_1  \label{(2,2)} \end{eqnarray}
where $e_{21} = e_2 - e_1$.  We note that $e_1 = 1$, but leave the equations above in general form for future reference.

Two related problems are inherent in Equations \ref{(0,0)} to \ref{(2,2)}.  First, all terms in the diagonal sum rules are positive definite, so Equation \ref{(2,2)} cannot hold for any values of the parameters.  We approach this issue in two ways.  First, we can simply ignore this equation and mathematically solve the remaining ones under the assumption that the well-behaved relationships between the parameters represents physical reality.  However, the fact that one equation is self-inconsistent brings into question the validity of truncating the sum-over-states expression to three states.  Alternatively, we can add an additional state to make all the three-level sums rules internally consistent, but then the sum rule $(3,3)$ will be self inconsistent.  Using this process {\em ad infinitum}, no matter how many states are added, there will always be an issue with the highest energy state, $N$, where the sum rule $(N,N)$ will be afflicted with the inconsistency pathology.

The second inherent problem is that the transition moments and energies that give a good approximation to the hyperpolarizability may not obey the sum rules.  In fact, it would be surprising if the same set of parameters would both obey the sum rules and lead to a good approximation of the nonlinear susceptibility since the equations scale differently with energy; that is, terms that are small in the SOS expression due to a large energy denominator will be large in the sum rules due to the multiplicative energy factor.  This seeming paradox can be resolved by recognizing that higher energy eigenstates can contribute to the sum rules without contributing to the nonlinear susceptibility, which is often found to be true,\cite{shafe11.02} but admittedly may not be generally true.

The question still remains if three states are appropriate for calculating limits.  The original strategy was based on the observation that the two-level model rigorously leads to a limit of the polarizability, $\alpha$, for a two state system,\cite{kuzyk13.01} and that adding more states causes $\alpha$ to decrease.  Furthermore, the two-level model is an exact description of a quantum system if $\alpha$ is at the limit.  With some modifications, aspects of these properties will be carried over to the case of $\beta$ and higher order nonlinearity.

Applying this argument directly to $\beta$ leads to the wrong prediction, that is, $\beta = 0$.  In effect, the two-level model holds only for systems with {\em minimum} hyperpolarizability.  The source of the issue lies in the two-level form of sum rule Equation \ref{(0,1)}, which demands that the dipole moment difference vanish when the transition moment between the ground and excited state is nonzero.  Conversely, the the transition moment must vanish when the dipole moment difference is nonzero.  This result is in accord with the linear harmonic oscillator,   which is centrosymmetric so has no dipole moment, and the transition moment to the first excited state is identically given by Equation \ref{xMAX}.  Furthermore, the transition moments to all other excited states vanish, making it an exact two-level system in which all sum rules are obeyed.  As such, the two-level model prediction is exact in this case.

Models of non-vanishing $\beta$ must therefore include at least 3 states, for which two independent dipolar terms contribute to the hyperpolarizability; one with a transition to the first excited state and the other to the second excited state.  The octupole term, on the other hand, mixes all three states, so vanishes in the two-level model.

Ironically, while the two-level model of $\beta$ cannot both obey the sum rules and give nonzero hyperpolarizability, it has been successful in modeling many dipolar molecules.  Similarly, the three-state model has been used to successfully model octupolar ones.\cite{Joffr92.01} However, since the two excited states in a system with octupolar symmetry are degenerate, an octupole is described in terms of only two state energies so is a two-level model of sorts. Clearly, the sum rules can be grossly violated in such a limited-state model, yet yield a good approximation for $\beta$.  Incidentally, a three-level model of the harmonic oscillator, which obeys the sum rules exactly, also gives $\beta = 0$.  This too is not surprising given the fact that the harmonic oscillator corresponds to a linear restoring force, so all nonlinear terms must vanish.

A two-state model is ideal for concentrating all the available strength of a transition into one state thus optimizing the polarizability, but at least 3 states are needed for the sum rules to be obeyed and for $\beta$ to be nonzero.  Paralleling the result for the polarizability, it was postulated that the minimum number of states is to be used that are simultaneously consistent with the sum rules, aside from the $(2,2)$ sum rule.  As such, the three-level model was chosen.  The two-state model of $\alpha$ is exact at the limit, so in analogy, the three-state model was postulated to be exact at the limit, though this does not preclude the three-level model from sometimes being a good approximation to $\beta$ when the system has a hyperpolarizability far short of the limit. This argument was proposed as an ansatz, i.e. a guess, which is applied to the calculation of the limits and the resulting predictions are compared with experiment as a test of its validity.

One might be tempted to apply all sum rule Equations \ref{(0,0)} to \ref{(1,2)} in the derivation of the limits.  However, as we will see below, this leads to a relationship between the energy spectrum and the transition moments.  Pick one energy spectrum and all other parameters become defined.  This clearly cannot be the case since the transition moments cannot be uniquely determined from the energy spectrum.  More colorfully stated, one cannot hear the shape of a drumhead.\cite{kac66.01}  The use of all three-level sum rules would thus eliminate many real systems.

Given that the three-level model is too restrictive if all the sum rules are used, a strategy must be developed to determine which sum rule(s) to eliminate.  In finite-state quantum systems, the sum rules that contain the highest-energy states are disobeyed by margins much greater than low-lying states.\cite{kuzyk06.01}  We can imagine the $(N,N)$ sum rule, which is self inconsistent, as an extreme case. The sum rules in the vicinity of this extreme, while self consistent, may be highly inaccurate; so, the simple criteria for an $N$-level system that we try here is,
\begin{equation}\label{sumRuleconstraint}
m+p < N .
\end{equation}
This constraint eliminates the sum rules that are nearest the truncation point under the assumption that $m+p$ characterizes the {\em order} of the sum rule.  In addition to culling out the inaccurate sum rules, this condition relaxes the number of constraints and allows two distinct quantum systems to share the same energy spectrum yet differ in their transition moments.

Equation \ref{sumRuleconstraint} directs us to ignore sum rules $(2,2)$ and $(1,2)$ when using the three-state model.  In principle, the excluded sum rules can be made more accurate by adding enough additional states, but if these additional states do not contribute to $\beta$, they can be ignored.  The trick is to find the minimum number of states needed to describe the nonlinearity (it appears to be three for the hyperpolarizability), and then eliminate enough of the sum rules so that the remaining ones are obeyed with enough degrees of freedom to span all possible quantum systems.  In the case of the three-state model, we proceed by ignoring sum rules $(2,2)$ and $(1,2)$.\cite{kuzyk00.01}

Since the dipole moments always appear as differences in the sum rules and in the SOS expressions used in calculating the hyperpolarizabilities,\cite{orr71.01} we reference all dipole moments to the ground state dipole moment, thus $\xi_{00} = 0$.  To be consistent with the original PRL paper,\cite{kuzyk00.01}, we define $E = e_2^{-1}$ and $x = \xi_{01}$.  Using this along with Equation \ref{xMAX}, sum rule (0,0) gives,
\begin{equation}\label{sum rul (0,0)}
\xi_{02} = a \sqrt{E \left( 1 - x^2 \right)},
\end{equation}
where $a = \pm 1$ ($a$ determines the sign of $\xi_{02}$).
Similarly, sum rule (1,1) gives,
\begin{equation}\label{sum rul (1,1)}
\xi_{12} = b \sqrt{\frac {E} {1-E} \left( 1 + x^2 \right)},
\end{equation}
where $b = \pm 1$ ($b$ is the sign of $\xi_{12}$) and the sum rule (0,1) with the help of Equations \ref{sum rul (0,0)} and \ref{sum rul (1,1)} gives,
\begin{equation}\label{sum rul (1,0)}
\xi_{11} = - ab \frac {2 - E} {\sqrt{1-E}} \frac {\sqrt{ 1 - x^4 }} {x}.
\end{equation}
Similarly, we can us sum rule (0, 2) to get
\begin{equation}\label{sum rul (2,0)}
\xi_{22} = a b \frac {1 - 2E} {\sqrt{1-E}} \cdot x \cdot \sqrt {\frac {1 + x^2} {1 - x^2}} .
\end{equation}
Note that we have used the fact that $b/a = ab$.

In calculating the fundamental limits of the first and second hyperpolarizabilities, Equations \ref{sum rul (0,0)} through \ref{sum rul (2,0)} are used to reexpress the sum-over-states\cite{orr71.01} expression in terms of $E$ and $x$ yielding,
\begin{equation}\label{3L-Ansatz-beta}
\beta = \beta_{max} f(E) G(x),
\end{equation}
where
\begin{equation}\label{beta-max-off-res}
\beta_{max} = \sqrt[4]{3} \left( \frac {e \hbar} {\sqrt{m}} \right)^3 \frac {N^{3/2}} {E_{10}^{7/2}} .
\end{equation}
The functions $f(E)$ and $G(x)$ are given by
\begin{equation}\label{DEFf(E)}
f(E) = (1-E)^{3/2} \left( E^2 + \frac {3} {2} E + 1 \right),
\end{equation}
and
\begin{equation}\label{defG(X)}
G(x) = \sqrt[4]{3} x \sqrt{\frac {3} {2} \left( 1 - x^4\right)},
\end{equation}
with
\begin{equation}\label{x/xMAX}
x= x_{10} / x_{10}^{max} \mbox{\vspace{1em} and \vspace{1em}} E = E_{10} / E_{20}.
\end{equation}

The maximum values of these functions are $G(\sqrt[-4]{3}) = 1$ and $F(0)=1$.  Assuming that $x$ and $E$ are independent, the hyperpolarizability is bounded by Equation \ref{beta-max-off-res}

The second hyperpolarizability expression does not decouple into the product of two functions as in Eq. \ref{3L-Ansatz-beta}, and the limit is given by,\cite{kuzyk00.01,kuzyk00.02}
\begin{equation}\label{gammamax}
-\left( \frac {e \hbar} {\sqrt{m}} \right)^4 \frac {N^2} {E_{10}^5} \leq \gamma \leq 4 \left( \frac {e \hbar} {\sqrt{m}} \right)^4  \frac {N^2} {E_{10}^5} \equiv \gamma_{max}.
\end{equation}
The negative and positive limits differ in their magnitudes by a factor of 4.

The transition moment matrix, normalized to $x_{10}^{max}$, and the energy eigenvalues, normalized to $E_{10}$, when the hyperpolarizability is near the limit is give by,
\begin{equation}
x = \left(
  \begin{array}{ccc}
    0.0 & 0.760 & 6.50 \times 10^{-6} \\
    0.760 & -2.15 & 1.26 \times 10^{-6} \\
    6.50 \times 10^{-6} & 1.26 \times 10^{-6}\times 10^5 & 1.47 \\
  \end{array}
\right), \hspace{.5em}
E = \left(
  \begin{array}{c}
    0 \\
    1 \\
    1 \times 10^{10} \\
  \end{array}
\right) .
\end{equation}
Note that we have used $E_{20}/E_{10} = 1 \times 10^{10}$ rather than the infinite value at the limit to avoid computational difficulties.  The difference between the calculated hyperpolarizability, using finite second excited state energy, and the limit is negligible.   The sum rule matrix and the RMS fluctuation of the sum rule matrix (defined later in Equation \ref{sigma}) near the limit are given by
\begin{equation}
SR = \left(
  \begin{array}{ccc}
    0.0 & 0.0 & 0.0 \\
    0.0 & 1.0 & 2.0 \times 10^5 \\
    0.0 & 2.0 \times 10^5 & -2.0 \\
  \end{array}
\right), \hspace{2em} \sigma = 9.54 \times 10^4 .
\end{equation}

The three-level ansatz used in calculating the limits leads to a $(1,2)$ sum rule that diverges at the limit when $E_{20} \rightarrow \infty$.  Recall that the $(1,2)$ sum rule is ignored to make $G(x)$ and $f(E)$ independent.  The limit given by Equation \ref{beta-max-off-res} is found to be obeyed for all quantum systems, so neglecting this sum rule when calculating the upper bound of $\beta$ seems appropriate.  The approach works despite the infinite violation of the sum rules near the truncation boundary because the infinite energy appears in the denominator when calculating the hyperpolarizability.  When approximating the hyperpolarizability using more states, the challenge is to balance the accuracy of the sum rules while making enough degrees of freedom available to span all possible quantum systems.  The three-level ansatz excluding sum rule $(1,2)$ seems to give nearly the right balance.

\section{Discussion}

\subsection{Three-State Model}\label{3L-model}

In the previous section, $x$ and $E$ were argued to be independent variables within the three-state manifold due to the fact that sum rule $(1,2)$ is most likely unreliable based on the observation that the $(m,p)$ sum rule for an $N$-state system is less accurate when $m+p \geq N$, becoming highly inaccurate as $m+p \rightarrow 2N$.  It must be stressed that the sum rules are exact and cannot be violated when an infinite number of states is included in the sum.  The apparent violation in real systems is due solely to truncation of the sum, where the missing terms in the sum are responsible.

The calculations of the limits leading to Equation \ref{beta-max-off-res} assume that a system with a hyperpolarizability at the limit has only three-states that contribute to the hyperpolarizability.  This conjecture is called the three-level ansatz.  As we saw in the previous section, a heuristic argument, guided by experimental observations, was used to deal with pathologies.

Having three states contribute to the hyperpolarizability is not equivalent to the system being a three-state system, i.e. that only three energy eigenstates exist.  Based on the fact that these other states may contribute to the sum rules even though they do not contribute to the hyperpolarizability motivates the assumption that $x$ and $E$ are independent.

Next we consider the consequences of including sun rule $(1,2)$, which we will show leads to a relationship between $x$ and $E$.  Substituting Equations \ref{sum rul (0,0)} through \ref{sum rul (2,0)} into sum rule $(1,2)$ in Equation \ref{(1,2)} yields, after considerable algebra
\begin{equation}\label{quadratic E&x}
-3 E x^2 + 2 - E =0,
\end{equation}
which can be solved for E(x),
\begin{equation}\label{E(x)}
E(x) = \frac {2} {3 x^2 + 1},
\end{equation}
or for x(E),
\begin{equation}\label{x(E)}
x(E) = \sqrt{ \frac {2 - E} {3E}}.
\end{equation}
Since $E$ and $X$ by definition cannot exceed unity, Equations \ref{E(x)} and \ref{x(E)} demand that
\begin{equation}\label{constraint}
\frac {1} {2} \leq E \leq 1 \mbox{\hspace{1em} and \hspace{1em}} \frac {1} {\sqrt{3}} \leq \left| x \right| \leq 1 .
\end{equation}

\begin{figure}
\centering
\includegraphics{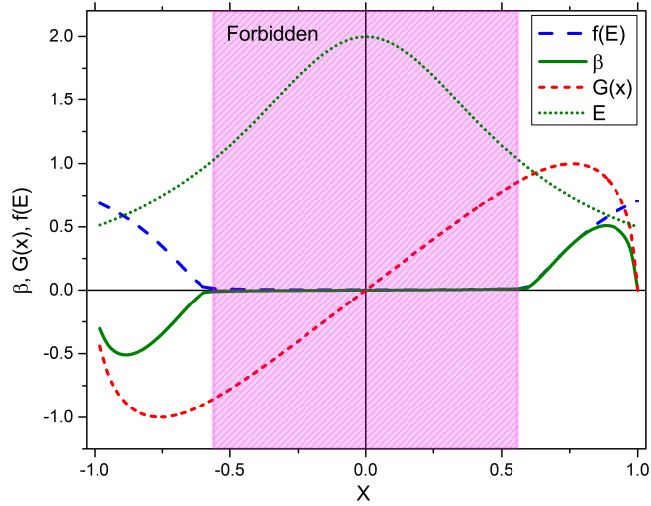}
\caption{$\beta$, $G(x)$, and $f(E)$ as a function of $x$.  The shaded region is forbidden because $E > 1$.}
\label{Fig:f(E(x))}
\end{figure}

\begin{figure}
\centering
\includegraphics{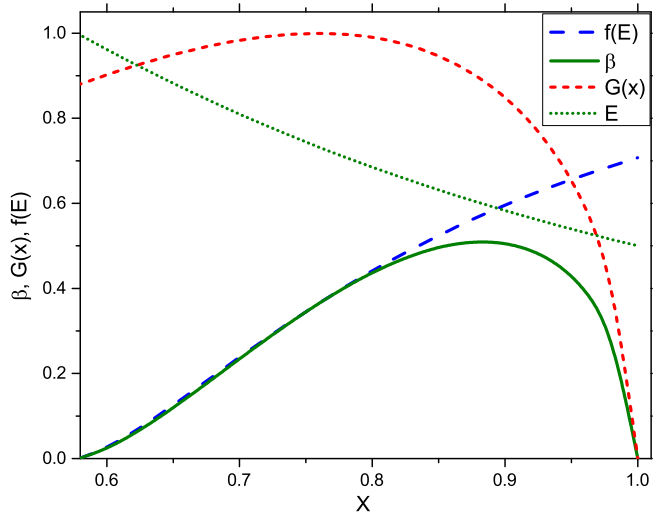}
\caption{Closeup of peak hyperpolarizability region.}
\label{Fig:f(E(x))1}
\end{figure}

Inequalities \ref{constraint} epitomize the problems associated with using sum rule $(1,2)$ -- a domain of legitimate parameters that are found in real quantum systems is excluded.  Even so, it is instructive to follow through with the process.  Figure \ref{Fig:f(E(x))} shows a plot of $f(x) = f(E(x))$, whose x-dependence is calculated from Equation \ref{E(x)}; $G(x)$; and, $\beta(x) = f(E(x)) G(x)$.  The constraints given by Inequalities \ref{constraint} exclude the shaded region. An expanded view of the peak is shown in Figure \ref{Fig:f(E(x))1}

The hyperpolarizability peaks at $\beta = 0.5109$ for $x = 0.8838$ and $E = 0.5982$.  Thus, the upper bound of the hyperpolarizability is underestimated with respect to potential optimization studies, which give $\beta = 0.7089$\cite{zhou06.01,zhou07.02} and Monte Carlo studies, which give $\beta = 1$\cite{kuzyk08.01,shafe10.01,shafe11.01}.

\subsection{The Four-state model}\label{sec:FourState}

In Section \ref{3L-model}, the three-state model suggested that Equation \ref{sumRuleconstraint} is a reasonable constraint that determines which sum rules are to be used to ensure that the domain is not unduly restricted.  It is worthwhile to consider the number of sum rule equations that should be used for an $N$-state system.  The number of independent sum rule equations obeying Equation \ref{sumRuleconstraint} is given by:
\begin{equation}\label{Constrained-Num-sum-Rules}
  N_c = \left\{
          \begin{array}{c}
            \frac {N(N+2)} {4}, \hspace{1em} N \hspace{1em} \mbox{even}\\
            \frac {(N+1)^2} {4}, \hspace{1em}N \hspace{1em} \mbox{odd}\\
          \end{array}
        \right. .
\end{equation}

Since only dipole moment differences appear in all sum rule expressions and nonlinear susceptibilities, we can without loss of generality make the choice $\xi_{00} = 0$.  This leaves,
\begin{equation}\label{N-dipole}
  N_d = \frac {(N+2)(N-1)} {2}
\end{equation}
independent dipole moment matrix elements.  Given that only energy differences appear in the sum rules and the SOS expression of the hyperpolarizabilities, we set $E_0 = 0$.  Furthermore, the energy $E_{10}$ defines the energy scale to which all others are normalized when calculating the intrinsic properties, so the number of independent energies is $N-2$.  Thus, with $N_c$ sum rule equations, the total number of free parameters is $N_f = N_d + (N-2) - N_c$, or
\begin{equation}\label{Free-Parameters}
  N_f = \left\{
          \begin{array}{c}
            \frac {(N+6)(N-2)} {4}, \hspace{1em} N \hspace{1em} \mbox{even}\\
            \frac {(N+6)(N-2) - 1} {4}, \hspace{1em}N \hspace{1em} \mbox{odd}\\
          \end{array}
        \right. .
\end{equation}

Equation \ref{Free-Parameters} for $N=3$ yields $N_f = 2$.  The two parameters that we chose in Section \ref{3L-model} were $x$ and $E$.  For $N=4$, $N_f = 5$, so we have the freedom to express the energies and tranasition moments in terms of any 5 parameters.  The sum rules for a four level system are given by,
\begin{eqnarray}\label{4L sum rules}
  (0,0)&:&  x^2 E + y^2 F + z^2 = E \nonumber \\
  (0,1)&:&  x \xi_{11} E + y \xi_{12} (2F-E) + z \xi_{13} (2-E) = 0 \nonumber \\
  (0,2)&:&  x \xi_{12} (2E-F) + y \xi_{22}F + z \xi_{23} (2-F) = 0 \nonumber \\
  (1,1)&:&  - x^2 E +  \xi_{12}^2 (F-E)+ \xi_{13}^2 (1-E)  = E \nonumber \\
  (0,3)&:&  x \xi_{13} (2E-1) + y \xi_{23} (2F-1) + z \xi_{33} = 0 \nonumber \\
  (1,2)&:&  \left( \xi_{22} - \xi_{11} \right) \xi_{12} (F-E) + \xi_{13}\xi_{23} (2 - E - F) = x y (E+F) \nonumber \\
  (1,3)&:&  \left( \xi_{33} - \xi_{11} \right) \xi_{13} (1-E) + \xi_{12}\xi_{23} (2F - E - 1) = x z (E+1) \nonumber \\
  (2,2)&:&  - y^2 F - \xi_{12}^2 (F-E)+ \xi_{23}^2 (1-F)  = E \nonumber \\
  (2,3)&:&  \left( \xi_{33} - \xi_{22} \right) \xi_{23} (1-F) + \xi_{12}\xi_{13} (2E - F - 1) = y z (1+F) , \nonumber \\
\end{eqnarray}
where $E = E_{10} / E_{30}$, $F = E_{20} / E_{30}$, $x = \xi_{01}$, $y = \xi_{02}$, and $z = \xi_{03}$. Equation \ref{sumRuleconstraint} demands that we use only the first six sum rules in Equation \ref{4L sum rules}.

The approach to expressing the moments in terms of a reduced number of parameters using the sum rules is as follows.  We choose as the free parameters $E $, $F$, $x$, $y$, and $\xi_{13}$.  The energies, by virtue of their assumed ordering, must obey,
\begin{equation}\label{eq:cond1}
E \leq F \leq 1 ;
\end{equation}
and, the transition moments according to sum rule $(0,0)$ must obey
\begin{equation}\label{eq:cond2}
x^2 \leq 1
\end{equation}
and
\begin{equation}\label{eq:cond3}
y^2 \leq (1 - x^2) E/F .
\end{equation}
Finally, sum rule $(1,1)$ demands that
\begin{equation}\label{eq:cond4}
\xi_{13}^2 \leq E (1 + x^2)/(1-E) .
\end{equation}

The system of equations can be solved by starting with the $(0,0)$ sum rule, which gives,
\begin{equation}\label{z()}
z(E,F,x,y,\xi_{13}) = a \sqrt{ \left(1 - x^2 \right) E - y^2 F} ,
\end{equation}
where $a = \pm 1$, and determines the sign of $z$.  Sum rule $(1,1)$ can then be used to solve for $\xi_{12}$,
\begin{equation}\label{xi12()}
\xi_{12}(E,F,x,y,\xi_{13}) = b \sqrt{ \frac {E \left( 1+x^2 \right) - \xi_{13}^2 (1-E)} {F-E} },
\end{equation}
where $b = \pm 1$ determines the sign of $\xi_{12}$.  Sum rule $(0,1)$ can then be used to get $\xi_{11}$,
\begin{equation}\label{xi11()}
\xi_{11}(E,F,x,y,\xi_{13}) = \frac { \xi_{13} (E-2) z(\bullet) + y (E-2F) \xi_{12}(\bullet) } {xE},
\end{equation}
where we have used ``$\bullet$" as a shorthand to represent ``$E,F,x,y,\xi_{13}$".

To get $\xi_{22}$, we can can eliminate $\xi_{23}$ from sum rules $(0,2)$ and $(1,2)$, which yields,
\begin{eqnarray}\label{xi_22()}
\xi_{22}(E,F,x,y,\xi_{13}) &=& \frac {1} {z(\bullet) \xi_{12}(\bullet) (2-F)(F-E) - \xi_{13} y F (2-E-F)}  \nonumber \\
&\times& \left\{ xyz(\bullet)(E+F)(2-F)  \right. \nonumber \\
&&+ \xi_{11}(\bullet) \xi_{12}(\bullet) z(\bullet) (F-E)(2-F) \nonumber \\
&&+ \left. x \xi_{13} \xi_{12}(\bullet) (2E-F)(2-E-F) \right\} .
\end{eqnarray}
Then using sum rule $(0,2)$ and Equation \ref{xi_22()}, we can solve for $\xi_{23}$,
\begin{equation}\label{xi23()}
\xi_{23}(E,F,x,y,\xi_{13}) = \frac {x \xi_{12}(\bullet) (F-2E) - y F \xi_{22}(\bullet)} {z(\bullet)(2-F)} .
\end{equation}
Finally, we get $\xi_{33}$ from the $(0,3)$ sum rule,
\begin{equation}\label{xi_{33}()}
\xi_{33}(E,F,x,y,\xi_{13}) = \frac {x \xi_{13}(1-2E) + y \xi_{23}(\bullet) (1-2F)} {z(\bullet)} .
\end{equation}

To apply the above approach, the parameters $E$, $F$, $x$, $y$, and $\xi_{13}$ are selected according to the constraints described above, and the rest of the parameters can be determined from Equations \ref{z()} through \ref{xi_{33}()}.

\subsection{Results}

\begin{figure}
\centering
\includegraphics{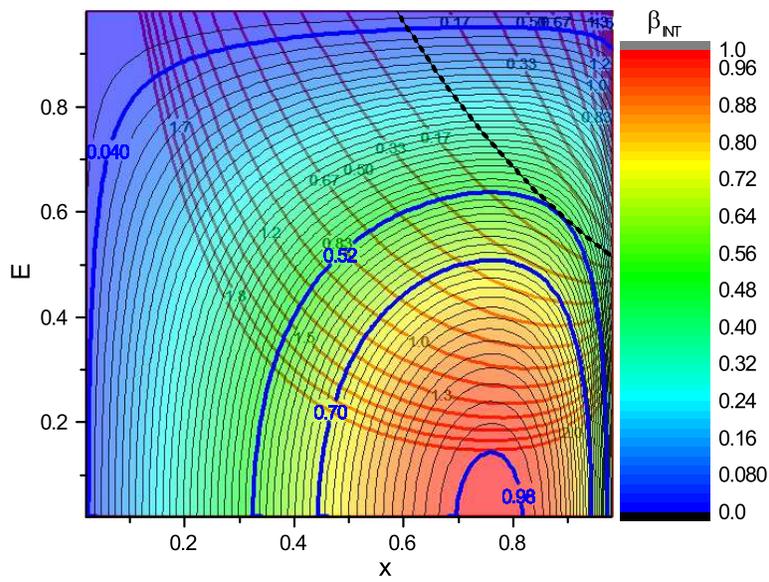}
\caption{Contour plot of $\beta_{INT}$ as a function of $E$ and $x$ in a three-level model where sum rule (1,2) is not used.  $\beta_{INT} = 0.98$, $\beta_{INT} = 0.7$, $\beta_{INT} = 0.52$, and $\beta_{INT} = 0.04$ are shown as solid thick blue contours.  Overlayed on the plot are the contours of the root mean square deviation, $\sigma$, from the sum rules.  The dark dashed curve highlights $\sigma = 0$.}
\label{Fig:BetaSigma}
\end{figure}
The above analysis can be used to study the properties of the three- and four-level models near the fundamental limit and to test the convergence of the sum rules.  Defining the intrinsic hyperpolarizability, $\beta_{INT}$ as
\begin{equation}\label{betaINT}
\beta_{INT} = \frac {\beta} {\beta_{MAX}} = f(E) G(x),
\end{equation}
Figure \ref{Fig:BetaSigma} shows a contour plot of $\beta_{INT}$ as a function of $E$ and $x$, which assumes that these two parameters are independent.  $\beta_{INT} = 0.98$, $\beta_{INT} = 0.7$, $\beta_{INT} = 0.52$, and $\beta_{INT} = 0.04$ are shown as dark (blue) contours.  Overlayed on the plot are the contours of the mean square deviation from the sum rules, $\sigma$, defined as
\begin{equation}\label{sigma}
\sigma = \frac {\sqrt{ \sum_{i = 1}^N \sum_{j = 1}^N \left( SR_{i,j} - \delta_{ij} \right)^2}} {N} ,
\end{equation}
where  $SR_{i,j}$ is given by
\begin{equation}\label{Deviation}
SR_{i,j} = \sum_{n=0}^\infty \xi_{in} \xi_{nj} \left( e_n - \frac {1} {2} \left[ e_i + e_j \right] \right).
\end{equation}
The thick black dashed curve highlights $\sigma = 0$.

The contour represented by $\beta_{INT} = 0.7$ is near the maximum hyperpolarizability observed for solutions of the schrodinger equation for one-dimensional systems, which is found to be about $\beta_{INT} = 0.7088$.  At this value, the smallest rms sum rule deviation is about $\sigma = 0.37$.\cite{zhou06.01,zhou07.02}   The limit is given by $\beta_{INT} = 1$ where $\sigma$ diverges.  In contrast, when the hyperpolarizability is a tad from the limit with $\beta_{INT} = 0.99964$, then $\sigma = 6.56$.  Note that the largest value of the hyperpolarizability on the contour representing $\sigma = 0$ is $\beta_{INT} = 0.5109$.

There is nothing unique about any of these points on the plot: $\sigma$ is large when $\beta_{INT} = 1$ and when $\beta_{INT} = 0$ and the contour representing the largest calculated hyperpolarizability given by $\beta_{INT} = 0.7088$ is similar to all others.  Furthermore demanding that all possible sum rule equations are obeyed, as represented by the dashed contour with $\sigma = 0$, leads to a maximum value that is below what is observed.

\begin{figure}
\centering
\includegraphics{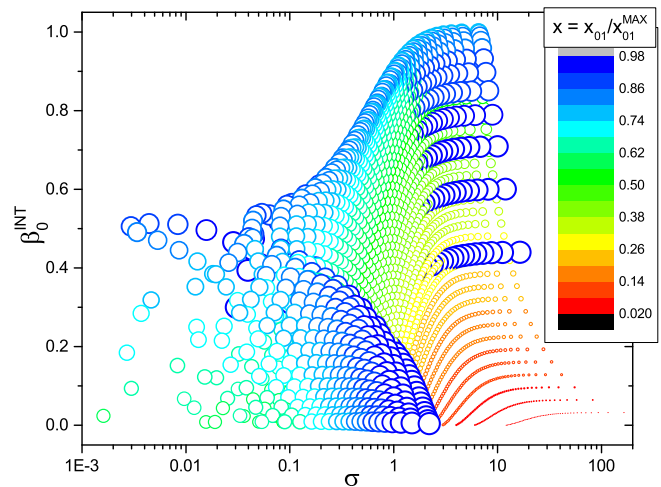}
\caption{The intrinsic hyperpolarizability as a function of $\sigma$.  The size (and colors) of the points are in proportion to the normalized transition moment.}
\label{Fig:BetaSigmaX}
\end{figure}

\begin{figure}
\centering
\includegraphics{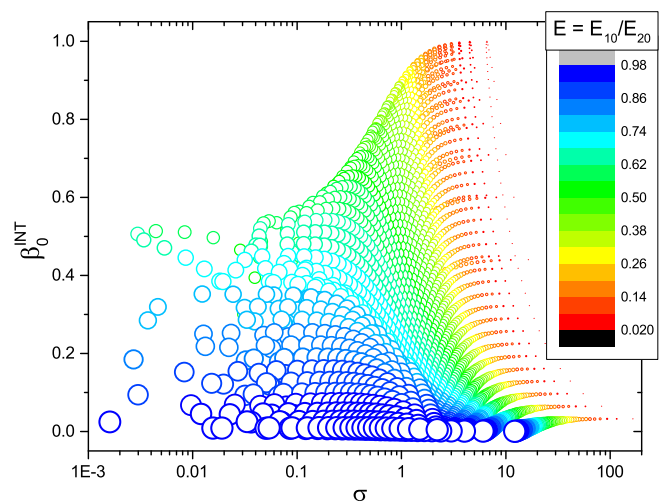}
\caption{The intrinsic hyperpolarizability as a function of $\sigma$.  The size (and colors) of the points are in proportion to the normalized energy.}
\label{Fig:BetaSigmaE}
\end{figure}

Figures \ref{Fig:BetaSigmaX} and \ref{Fig:BetaSigmaE} show a plot of the intrinsic hyperpolarizability as a function of $\sigma$ for $E$ and $X$ ranging from $0$ to $1$ in $0.02$ increments.  The size (and color) of the points in these two figures represent the values of $x$ and $E$ as labelled in the graph legends.  These plots represent the same data as in Figure \ref{Fig:BetaSigma}, but better show the range in hyperpolarizability.

Several conclusions can be drawn from this data.  First, the hyperpolarizability spans from zero to some maximum value determined by how much the sum rules are disobeyed.  If the sum rules are fully obeyed, the largest possible value is $\beta_{INT} = 0.5109$ and $\beta_{INT} = 1$ if the three-state sum rules diverge from compliance.  The important conclusion here is that the three-state model for the hyperpolarizability, if required to obey all the sum rules (except for $(2,2)$ of course, which never holds), leads to a limit that falls short of what is observed in one-dimensional quantum systems.  If the $(1,2)$ sum rule is excluded from consideration, one gets a limit that is a tad higher than what is observed.

Next we consider the four-level model in terms of the 5 parameters as described in Section \ref{sec:FourState}.  Restricting the space to five parameters may seem more restrictive than the Monte Carlo technique, which uses only the diagonal sum rules $(0,0)$, $(1,1)$ and $(2,2)$ -- a total of three constraining equations. In contrast, using the condition that $m+p<N$, the sum rules used are $(0,0)$, $(1,1)$, $(0,1)$, $(0,2)$, $(0,3)$, and $(1,2)$.  The latter approach uses the antidiagonal of the sum rule matrix as the demarcation that selects the sum rules, which in the three-state model yields the sum rules used in calculating the limit.

\begin{figure}
\centering
\includegraphics{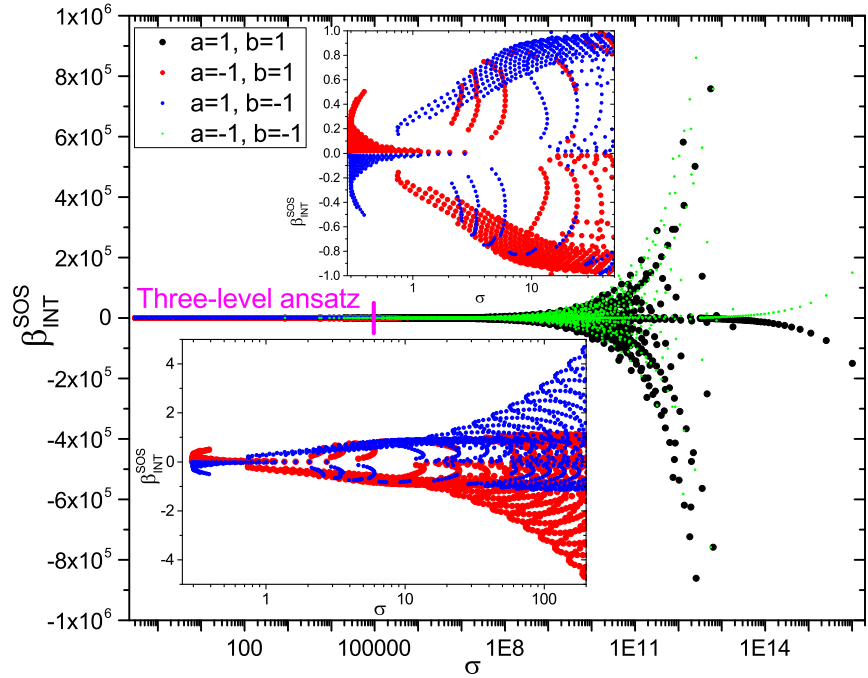}
\caption{The intrinsic hyperpolarizability as a function of the root mean square deviation of the sum rules derived by sampling the independent parameters $E$, $F$, $x$, $y$, and $\xi_{13}$ over their domains in increments of 0.02 in a raster scan pattern that is forced to obey the conditions imposed on the parameters by Equation \ref{eq:cond1} to \ref{eq:cond4}.  The insets show magnified views.}
\label{fig:AntiDiag}
\end{figure}
We sample independent parameters $E$, $F$, $x$, $y$, and $\xi_{13}$ over their domain in increments of 0.02 (i.e. $x = 0, 0.2, 0.4, \dots, 1.0$, etc.) in a raster scan pattern that is forced to obey the conditions imposed on the parameters by Equation \ref{eq:cond1} to \ref{eq:cond4}.  Figure \ref{fig:AntiDiag} shows the intrinsic hyperpolarizability as a function of the root mean square deviation of the sum rules for the full data set.  The insets show magnified views of various parts of the plots, and the color coding (and size) of the points distinguishes between the signs of the matrix elements as defined by the parameters $a$ and $b$ in Equations \ref{z()} and \ref{xi12()}.  For reference, the vertical line on the $\sigma$-axis labels the root mean square deviation of the sum rules for the three-level ansatz when $E_{20} = 10^{10} E_{10}$.

Within the sensitivity of the 0.02 sampling window, the intrinsic hyperpolarizability of the four-level model is less than the calculated limit of unity for root mean square deviations of the sum rules below about 15.  However, while all the sum rules below the antidiagonal are exactly obeyed, the full set of sum rules are grossly disobeyed.  The largest values observed at this resolution of sampling are almost a factor of $10^6$ above the limit and well above all real systems studied to date.  This implies that the anti-diagonal constraint is not restrictive enough to lead to a limit.

The Monte Carlo technique and the antidiagonal approach are two limiting cases of constraints, with the latter technique avoiding sum rules that include states near the truncation level.  The Monte Carlo approach, which uses only the diagonal sum rules, gives hyperpolarizabilities that are bounded by the same value as the limit derived from the three-level ansatz.  In contrast, the antidiagonal sum rules for a four-state system yields hyperpolarizabilities that greatly exceed unity.

It would appear that the diagonal sum rules are the key ingredient in getting the same limit as with the three-level ansatz.  The observation that the three-state model constrained by the antidiagonal sum rules gives the same limit as Monte Carlo sampling using an arbitrary number of states is an incredible result.  Since Monte Carlo calculations show that the three-level model is indeed bounded by unity, it would appear that the $(1,2)$ sum rule is not needed to get the same limit as with the standard approach -- though relaxing the $(1,2)$ sum rule gives more degrees of freedom.

Clearly, the three-state system is unique in that the antidiagonal demarcation constraint includes all diagonal sum rules, namely $(0,0)$ and $(1,1)$.  Motivated by these observations, we introduce a hybrid sampling technique that uses the antidiagonal demarcation constraint, but includes also all of the diagonal sum rules.  For the four-state model, this adds only one additional constraint, namely the $(2,2)$ sum rule.  Rather than adding this constraint algebraically, an approach which is already messy, we implement the raster sampling technique, but eliminate all points that violate the $(2,2)$ sum rule within a given tolerance level.

\begin{figure}
\centering
\includegraphics{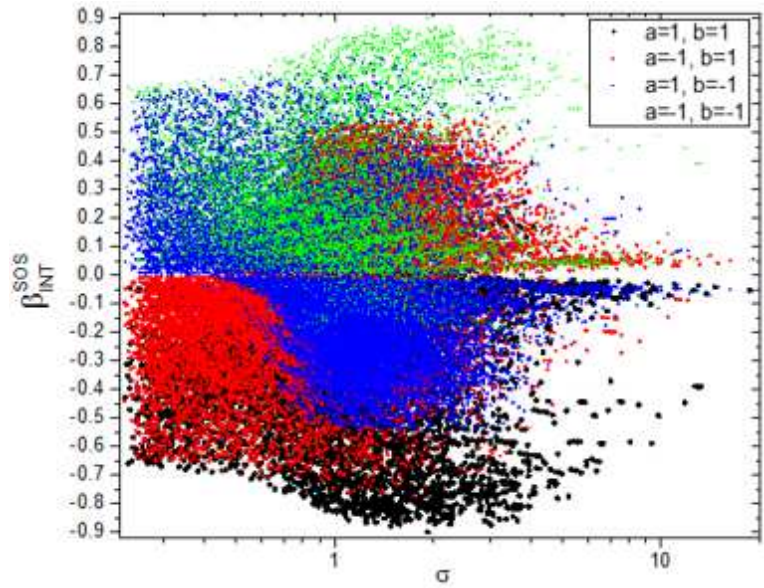}
\caption{The intrinsic hyperpolarizability as a function of the root mean square deviation of the sum rules derived by sampling the independent parameters $E$, $F$, $x$, $y$, and $\xi_{13}$ over their domains in increments of 0.05 in a raster scan pattern that is forced to obey the conditions imposed on the parameters by Equation \ref{eq:cond1} to \ref{eq:cond4}.  Points in which the $(2,2)$ sum rule is violated by more than 10\% are eliminated.}
\label{fig:AntiDiag1}
\end{figure}
Figure \ref{fig:AntiDiag1} shows the result of hybrid sampling for the four-level model.  $a$ and $b$ represent the signs of the transition moments as described above. The most striking result is that the intrinsic hyperpolarizability falls below 1.0, as expected since off-diagonal sum rules are used in addition to the diagonal sum rules, which alone are known to yield unity.  Furthermore, the RMS deviation of the sum rules has dropped from $\sigma = 10^{16}$ down to $\sigma < 20$.  Note that noncompliance with the sum rules does not necessarily lead to a large hyperpolarizability as long as the hybrid sum rule constraint is met.

The largest intrinsic hyperpolarizability observed is a tad over 0.9.  However, given the course increment used in sampling, larger values of $\beta$ are most likely there but missed.  As a case in point, since the three-level model is a subset of this data, $\beta_{INT} = 1$ should appear in the plot; but, since energy sampling is in increments of 0.05, the smallest energy ratio is $E = 0.05$, thus missing $\beta_{INT} = 1$ and the associated divergence of  $\sigma$.

As more states are added, since the three-level case is always a subset, the limit will be given by at least unity; but with the extra degrees of freedom, there may be many more ways to get unity.  The Monte Carlo calculations show this behavior, where (1) the limit remains at unity as the number of states is increased; (2) more combinations of parameters give unity; and (3) by virtue of the larger number of degrees of freedom with increasing number of states, the proportion of samples near the limit relative to all possible hyperpolarizabilities becomes smaller.\cite{kuzyk08.01,shafe10.01,shafe11.01}  The last point is an important one with regards to why so many materials fall far short of the limit.  Exceptional ones are rare.

The hybrid sum rule constraints provide an interesting opportunity for constructing hypothetical systems in which the sum rules are exactly obeyed for all states that contribute to the hyperpolarizability.  As an example, the three-level ansatz is a deliberately contrived system in which all the sum rules within the hybrid constraints hold for all the states that contribute to the nonlinearity of the system.  If one could algebraically concoct a system of Equations analogous to Equations \ref{z()} to \ref{xi_{33}()} that also include all diagonal sum rules (excluding $(N,N)$) for arbitrarily large number of states, then a small subset of states that have zero dipole transitions to higher-energy states would obey all of the sum rules and would be the only states contributing to the hyperpolarizability.  This type of construction, if possible, would eliminate the pathologies that plague sum rules calculations that use a small number of states. This type of approach may be useful in finding the ever-elusive proof of the three-level ansatz, if indeed it is true.\cite{shafe13.01}

\begin{figure}
\centering
\includegraphics{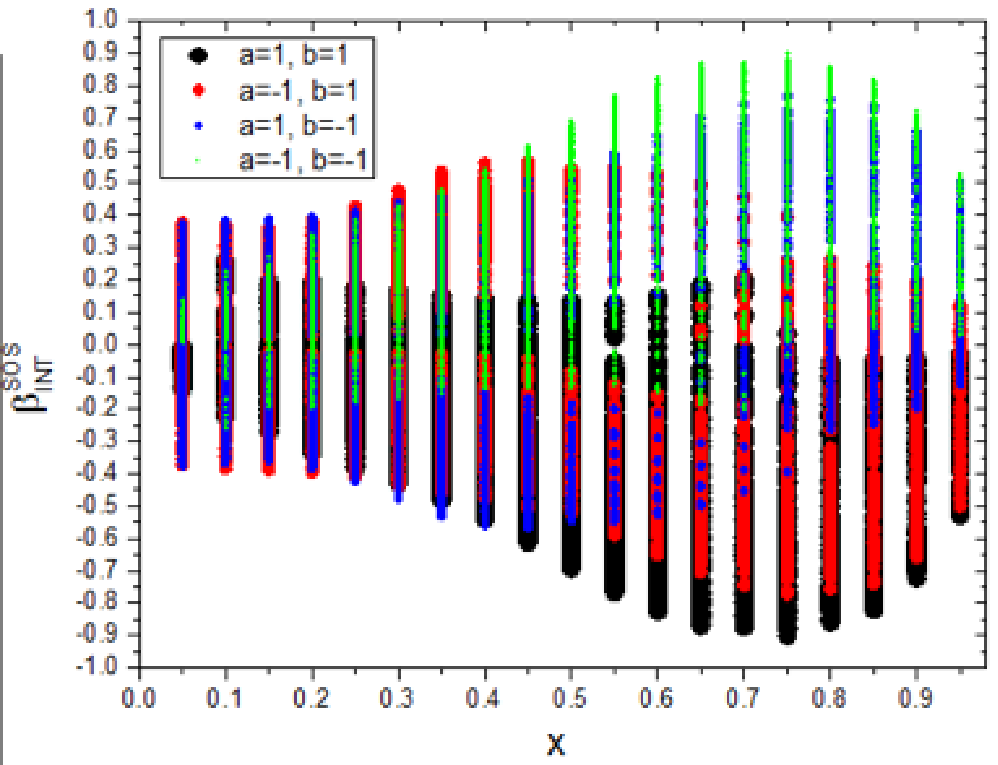}
\includegraphics{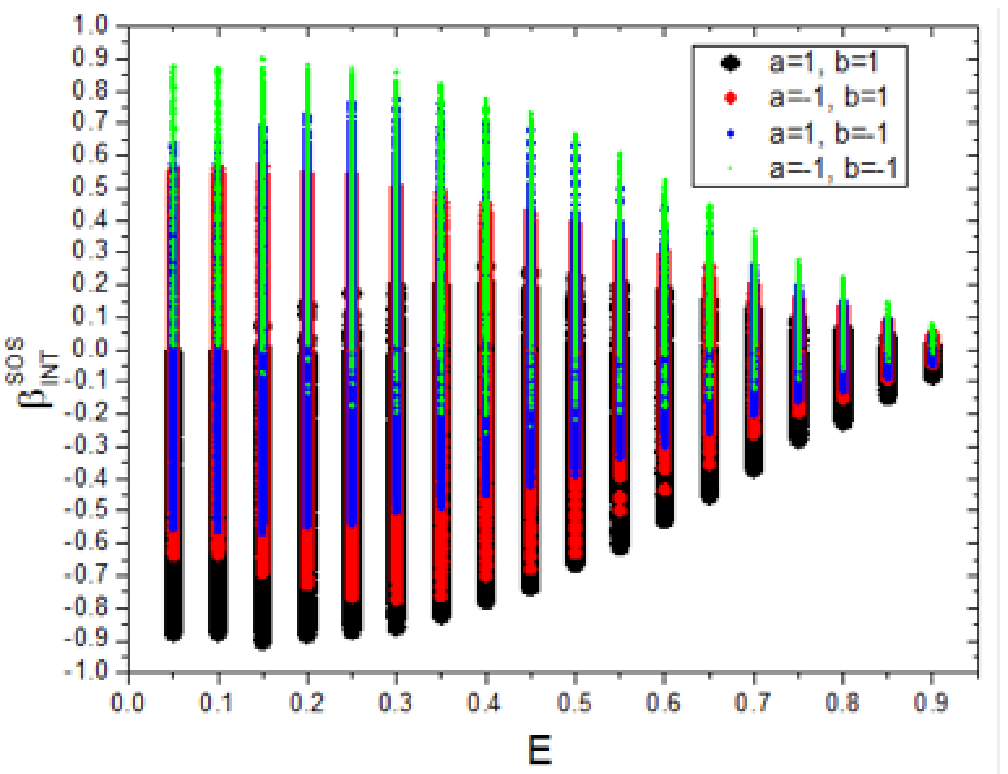}
\caption{The intrinsic hyperpolarizability as a function of $x$ (top) and $E$ (bottom) for all values of $F$, $y$, and $\xi_{13}$  in which the $(2,2)$ sum rule is violated by no more than 10\%.}
\label{fig:AntiDiagxE}
\end{figure}
Keeping in mind the coarseness of the sampling, we conclude with some general observations of the four-level model under the hybrid sum rule constraint.  Figure \ref{fig:AntiDiagxE} shows the intrinsic hyperpolarizability as a function of $x$ and $E$. The vertical range represents variations in all the other parameters and the color coding (thickness of points) represents the signs of the transition moments.  The outer envelope of each plot shows the extreme values of $\beta$.  Interestingly, the largest value of the intrinsic hyperpolarizability is for $x \approx 0.75$ and for energy $E$ near zero, both in agreement with the three-level model.  As such, the additional degrees of freedom in the four-level model do not appear to lead to larger $\beta$, nor can $\beta$ be larger in a four-state model with values of $x$ and $E$ other than those that optimize the three-level model with $x$ and $E$ being treated as independent.  The fact that $\beta_{INT}$ does not reach unity is due to the coarseness of the raster scan.

\begin{figure}
\centering
\includegraphics{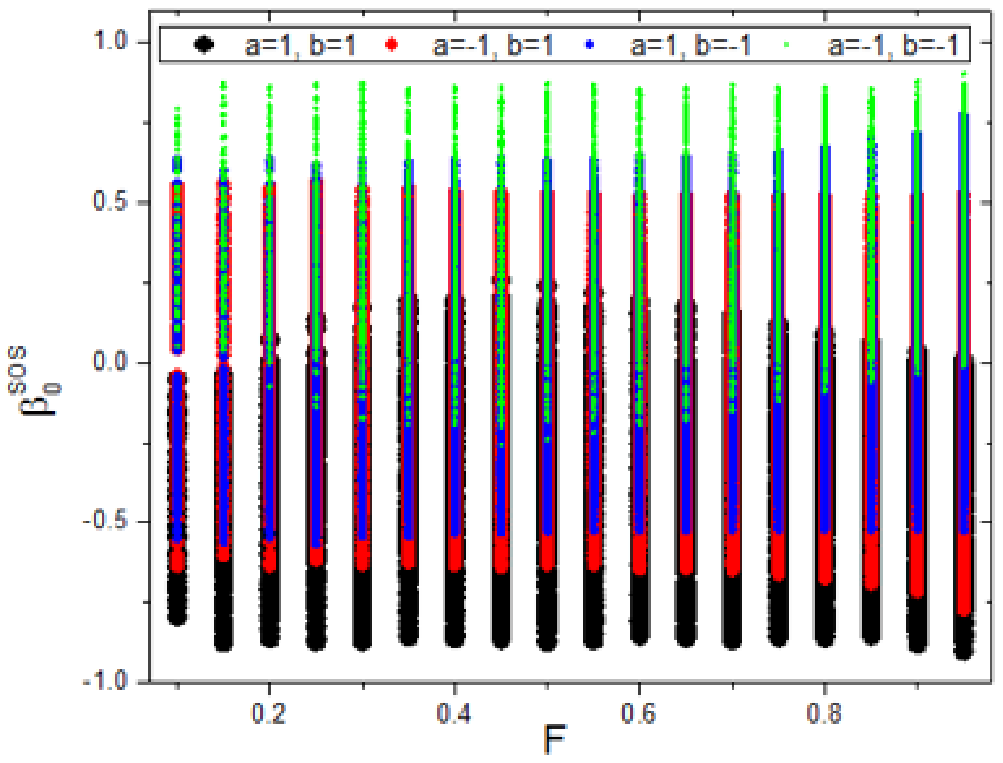}
\caption{The intrinsic hyperpolarizability as a function of $F$ when all other parameters are varied in a raster scan in 0.05 increments as in Figure \ref{fig:AntiDiagxE}.}
\label{fig:AntiDiagxF}
\end{figure}
Figure \ref{fig:AntiDiagxF} shows the hyperpolarizability as a function of $F$ when all other parameters are varied in a raster scan in 0.05 increments.  Within statistical noise, every value of $F$ is associated with a hyperpolarizability near the limit.  This suggests that the energy of the third excited state is irrelevant to $\beta_{INT}$.  Thus, the most critical parameter is $E$, which yields a hyperpolarizability at the limit when the second excited state energy is much larger than the first, and vanishes when the first two excited states are degenerate.

\begin{figure}
\centering
\includegraphics{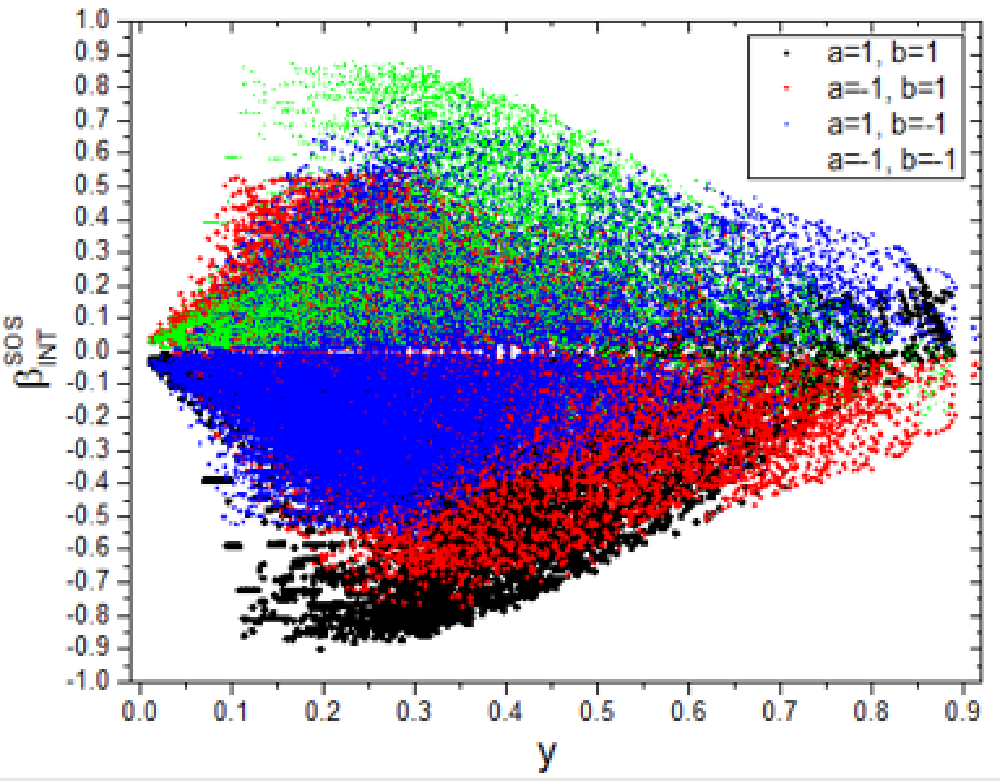}
\includegraphics{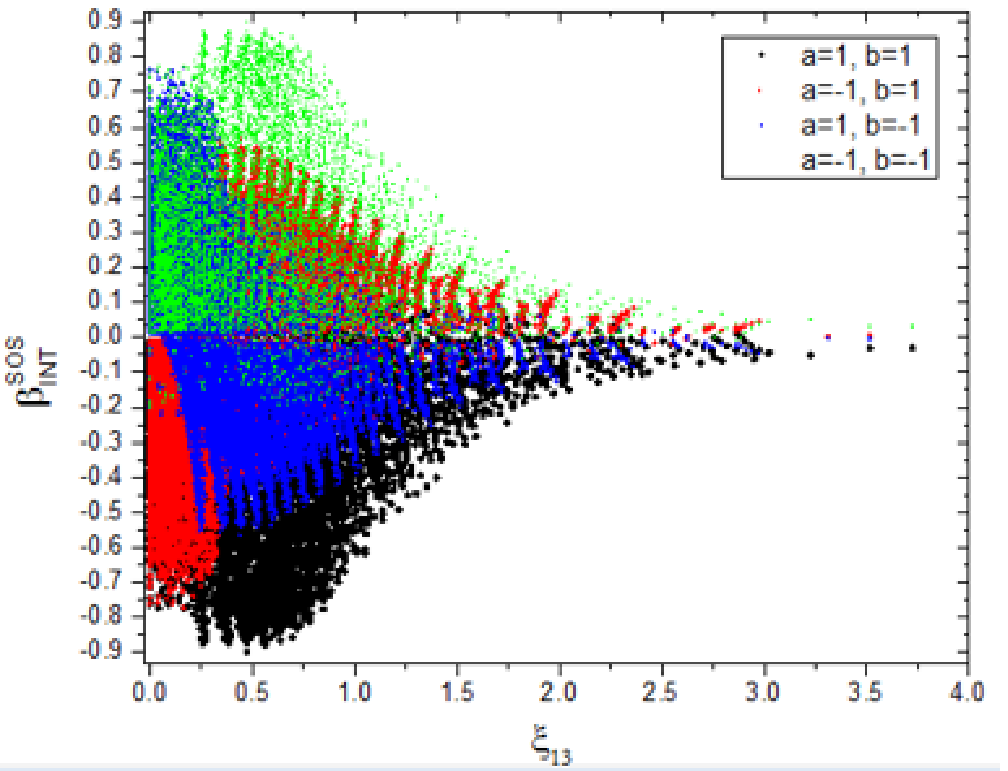}
\caption{The intrinsic hyperpolarizability as a function of $y$ (top) and $\xi_{13}$ (bottom) for all values of $F$, $y$, and $\xi_{13}$  in which the $(2,2)$ sum rule is violated by no more than 10\%.  The largest values of $\beta_{INT}$ are for small $\xi_{13}$ and $y$.  Note that the points near $y=0$ are sparse, but the envelope approaches unity at $y=0$ if the raster increment is made finer.}
\label{fig:AntiDiagy-xi13}
\end{figure}
Figure \ref{fig:AntiDiagy-xi13} shows a plot of the intrinsic hyperpolarizability as a function of $y$ and $\xi_{13}$.  The largest intrinsic hyperpolarizabilities are in the range when these two parameters are small, suggesting that the three-level limit may be the best.

\section{Conclusion}

The fundamental limit of the hyperpolarizability was originally calculated using the three-level model and the sum rules to reduce the number of parameters.  Implicit in the calculation is the three-level ansatz, a guess that only three states contribute at the limit.  Because sum rule $(1,2)$ is ignored, the hyperpolarizability can be expressed in terms of two parameters, which are usually taken to be $x$ and $E$.  In this formulation, the hyperpolarizability is given by the product $f(E)G(x)$.  Optimization of each function separately leads to the limits.  This approach leads to a reasonable limit that is consistent with all reported measurements and calculations of toy model quantum systems.  However, at the limit, the sum rule that is ignored is violated by an infinite degree.  This motivated an investigation of the implications of the approximations used in the traditional approach.

To bring the $(1,2)$ sum rule into compliance, we next used it as an additional constraint, which demands that $x$ and $E$ are related.  We algebraically solved, in closed form, for the functions $x(E)$ and $E(x)$ and find that these functions are not defined on certain domains that are allowable in real systems.  Furthermore, the upper bound for $\beta$ that is calculated in this model yields $\beta_{INT} = 0.511$, below the values observed for real systems.  This unphysical behavior suggests that using all the sum rules overconstrains the system.

The next step of the analysis generalizes the three-level model to four states, and the sum rules are used to reduce the number of parameters.  There is great latitude in selecting the number of parameters to use and which sum rules to keep.  The ideal choice balances the number of degrees of freedom, which needs to be sufficient to span all quantum systems, but constrained enough to make predictions that agree with observations.

In one approach to the generalization of the three-state model, the antidiagonal demarcation method was used to determine which sum rules to keep, which for an $N$-level system requires that only the sum rules $(n,m)$ be used under the condition that $N<n+m$.  This approach does not yield an upper bound for the hyperpolarizability, and a coarse sampling of the parameter space finds values that are a million times larger than the observed upper bounds.  Clearly, this method is not restrictive enough to yield physically meaningful results.

An analysis of the Monte Carlo approach shows that the diagonal sum rules alone yield the same limit as calculated in the traditional approach, with the exception of the many state catastrophe, which is not normally observed in Monte Carlo sampling because it requires a very special set of parameters that are considered unphysical.\cite{shafe13.01} In light of these facts, we propose a hybrid method for generalizing the 3-level calculations of the limits using the antidiagonal condition AND requiring that all diagonal sum rules be used.  This hybrid method in the three-level limit reduces to the traditional approach for calculating the limits and a raster scan of the parameters gives intrinsic hyperpolarizability values that are bounded by the traditional three-level model calculation.

If the hybrid method can be algebraically generalized to more than 4 states, and there is no reason to believe that it can't, then every such model contains as a subset of its parameter space the three-state model, which is bounded by $\beta_{INT} = 1$.  As such, adding more states can never decrease the upper bound.  Furthermore, if the pattern observed by coarse raster scanning of the parameters of the four-state model holds for higher-level models, the upper bounds may also not get any larger by adding more states.  This observation may be related to the fact that the three-level ansatz seems to hold even at a local maximum that can be far from the global one.  The fact that the sum rules may not hold at a local maximum is not surprising given the fact that large violations of the sum rules are commonly observed without affecting the hyperpolarizability, especially if the energy of the state that causes the violation is sufficiently large.

\vspace{1em}

I thank the National Science Foundation (EECS-1128076) for supporting this work.

%\bibliography{\bibs}
%\bibliographystyle{\bstfile}

\end{document}